\documentclass{jpp}

\usepackage{graphicx}
\usepackage{natbib}

\ifCUPmtlplainloaded \else
  \checkfont{eurm10}
  \iffontfound
    \IfFileExists{upmath.sty}
      {\typeout{^^JFound AMS Euler Roman fonts on the system,
                   using the 'upmath' package.^^J}%
       \usepackage{upmath}}
      {\typeout{^^JFound AMS Euler Roman fonts on the system, but you
                   dont seem to have the}%
      \typeout{'upmath' package installed. JPP.cls can take advantage
                 of these fonts, if you use 'upmath' package.^^J}%
      }
  \else
  \fi
\fi

\ifCUPmtlplainloaded \else
  \checkfont{msam10}
  \iffontfound
    \IfFileExists{amssymb.sty}
      {\typeout{^^JFound AMS Symbol fonts on the system, using the
                'amssymb' package.^^J}%
       \usepackage{amssymb}%
         \let\leq=\leqslant
         \let\geq=\geqslant
      }{}
  \fi
\fi

\ifCUPmtlplainloaded \else
  \IfFileExists{amsbsy.sty}
    {\typeout{^^JFound the 'amsbsy' package on the system, using it.^^J}%
     \usepackage{amsbsy}}
    {}
\fi

\newsavebox{\astrutbox}
\sbox{\astrutbox}{\rule[-5pt]{0pt}{20pt}}

\title[Stationary distribution functions by MaxEnt principle]{Stationary distribution functions for Tokamak-plasmas in the weak-collisional transport regime by MaxEnt principle}

\author[G. Sonnino {\it et Al.}]%
{Giorgio\ls SONNINO$^1$\thanks{Email address for correspondence: gsonnino@ulb.ac.be},\ls Philippe\ls PEETERS$^2$,\ls Alberto\ls SONNINO$^3$,\ls \\ Pasquale\ls NARDONE$^2$, \ls and\ls Gy\"{o}rgy \ls STEINBRECHER$^4$\ns}

\affiliation{$^1$Universit{\'e} Libre de Bruxelles (U.L.B.), Department of Theoretical Physics and Mathematics, Campus de la Plaine C.P. 231 - Bvd du Triomphe, 1050 Brussels - Belgium \\ \&\\
Royal Military School (RMS), Av. de la Renaissance 30, 1000 Brussels, Belgium\\
[\affilskip]
$^2$Universit{\'e} Libre de Bruxelles (U.L.B.), Department of Physics, Campus de la Plaine C.P. 231 - Bvd du Triomphe, 1050 Brussels - Belgium\\[\affilskip]
$^3$Universit{\'e} Catholique de Louvain (UCL), Ecole Polytechnique de Louvain (EPL), Rue Archim$\grave{\rm e}$de, 1 bte L6.11.01, 1348 Louvain-la-Neuve, Belgium\\[\affilskip]
$^{4}$University of Craiova, Faculty of Exact Sciences, Str.A.I.Cuza Street 13, Craiova-200585, Romania}

\pubyear{2010}
\volume{650}
\pagerange{119--126}
\begin{document}

\maketitle

\begin{abstract}
In previous works, we derived stationary density distribution functions (DDF) where the local equilibrium is determined by imposing the maximum entropy (MaxEnt) principle, under the scale invariance restrictions, and the minimum entropy production theorem. In this paper we demonstrate that it is possible to reobtain these DDF solely from the MaxEnt principle subject to suitable scale invariant restrictions in all the variables.
For the sake of concreteness, we analyze the example of ohmic, fully ionized, tokamak-plasmas, in the weak-collisional transport regime. In this case we show that it is possible to reinterpret the stationary distribution function in terms of the Prigogine distribution function where the logarithm of the DDF is directly linked to the entropy production of the plasma. This leads to the suggestive idea that also the stationary neoclassical distribution functions, for magnetically confined plasmas in the collisional transport regimes, may be derived solely by the MaxEnt principle.

\end{abstract}


\section{\textbf{Introduction}}
In recent works we determined the expression of a stationary density distribution function (DDF), previously proposed in literature, by statistical thermodynamics \citep{sonninoPRE1}-\citep{sonninoEPJD2}. This DDF is denoted by ${\mathcal F}^{\alpha R}$ (with $\alpha=e,i$ for electrons and ions, respectively). The stationary DDF has been determined in three steps. Firstly, we considered open thermodynamic systems close to a local equilibrium state obeying to Prigogine's statistical thermodynamics. Successively, we defined the local equilibrium state by adopting a minimal number of hypotheses : the minimum entropy production (MEP) theorem and the MaxEnt principle, under two scale invariance restrictions. Finally, we linked the Prigogine probability distribution function with the DDF, which is obtained by perturbing the local equilibrium state (LES). 

\noindent The gyrokinetic (GK) theory makes often use of an initial distribution function of guiding centers. In the GK simulations, as well as in the GK theory, this initial distribution function is usually taken as a reference density distribution function if it depends only on the invariants of motion and it evolves slowly from the local equilibrium state i.e., in such a way that the guiding centers remain confined for sufficiently long time. 

\noindent The aim of this work is to show that the DDF may be derived by using solely the MaxEnt principle, subject to suitable restrictions. For the sake of concreteness, we shall analyze Tokamak ohmic-plasmas in the weak-collisional transport regime. We shall see that, in this case, the obtained DDF may be re-interpreted in terms of Prigogine's distribution function where the entropy production of the plasma is proportional to the logarithm of the DDF. 

\noindent The paper is organized as follows. In Section \ref{MaxEnt}, we shall re-derive, very briefly, the DDF derived by non-equilibrium statistical mechanics, by using {\it solely} the MaxEnt principle, under suitable scale invariance restrictions, without using the MEP. Of course, from the physical point of view, the special choice of the scale invariance restrictions can be justified by evoking the MEP. Section \ref{estimation} treats tokamak-plasmas in the weak collisional transport regime. We shall see that it is possible to set the free parameters appearing in the DDF in such a way that the logarithm of this stationary distribution function coincides exactly with the entropy production of the plasma estimated by the neoclassical theory. This leads to the suggestive idea that also the stationary distribution function derived by neoclassical theory may the  be obtained by MaxEnt principle (under suitable restrictions). 
 
\section{Derivation of a Stationary Distribution Function, Recently Mentioned in Literature, by MaxEnt Principle}\label{MaxEnt}

In this Section we shall prove that the DDF derived by non-equilibrium statistical mechanics \citep{sonninoPRE1}-\citep{sonninoEPJD2} can easily be obtained from maximum entropy principle with suitable scale invariant restrictions in all the variables. 

\noindent In the case of an axisymmetric magnetically confined plasma, after having performed the guiding center transformation, the necessary variables for describing the system reduce to four independent variables \citep{balescu1}. These variables are defined as follows. One of these ones is the {\it poloidal magnetic flux}, $\psi$, and another variable is the {\it particle kinetic energy per unit mass}, $w$, defined as $w=(v_{\parallel}^{2}+v_{\perp }^{2})/2$ with $v_{\parallel }$ denoting the parallel component of particle's velocity (which may actually be parallel or antiparallel to the magnetic field), and $v_{\perp }$ the absolute value of the perpendicular velocity \citep{balescu1}. The remaining two variables are the \textit{toroidal angular moment} $P_{\phi }$ and the variable linked to the pitch angle, $\lambda $. These quantities are defined as (for a rigorous definition, see any standard textbook such as, for example, \citep{balescu2}) 
\begin{eqnarray}\label{ddf1}
&P_{\phi }&=\psi +\frac{B_{0}}{\Omega _{0c}}\frac{Fv_{\parallel }}{\mid B\mid}\\
&\lambda&\equiv\frac{\mu }{w}=\frac{\sin ^{2}\theta _{P}}{2\mid B\mid }\qquad 
\mathrm{with}\quad \mu =\frac{v_{\perp }^{2}}{2\mid B\mid }
\end{eqnarray}
\noindent Here $\Omega _{0c}$ is the \textit{cyclotron frequency} associated with the magnetic field along the magnetic axis, $B_0$. $\mid B\mid $, $F$ and $\theta _{P}$ denote the {\it magnetic field intensity}, the {\it characteristic of axisymmetric toroidal field} depending on the {\it poloidal magnetic flux} $\psi$ and the \textit{pitch angle}, respectively. 

\noindent The expression for the reference (density of) distribution function ${\mathcal{F}}^{R}$ reads\footnote{We mention that this DDF has been previously proposed in Ref.~\citep{CdT} with $\Delta\lambda_1\rightarrow\infty$. The DDF in Ref.~\citep{CdT} has been introduced on the base of the Ion Cyclotron Radiation Heating (ICRH) FAST-plasma, simulated by using the Hybrid Magnetohydrodynamic Gyro-kinetic Code (HMGC) \citep{cardinali} and \citep{pizzuto}. However, the distribution function in Ref.~\citep{CdT} should be understood only as an {\it ad hoc} expression, which has been introduced for fitting the steady state equilibria in several physical scenarios by five free parameters devoid, at that level, of any physical interpretation. Hence, apart the presence of parameter $\Delta\lambda_1$, the major difference between the works in \citep{sonninoPRE1}-\citep{sonninoEPJD2} and the paper in Ref.~\citep{CdT} stands in the derivation method; based in the former works on a rigorous statistical mechanics approach.}
\begin{equation}\label{ddf7}
{\mathcal{F}}^{R}={\mathcal{N}}_{0}\!\left[  \frac{w}{\Theta}\right]
^{\gamma-1}\!\!\!\!\!\!\exp[-w/\Theta]\exp\left[  -\left( \frac{P_{\phi
}-P_{\phi0}}{\Delta P_{\phi}}\right)  ^{2}\right]  \exp\left[  -\left( 
\frac{\Delta\lambda_{0}}{\Delta\lambda_{1}}+\frac{w}{\Theta}\right)
\frac{(\lambda-\lambda_{0})^2}{\Delta\lambda_{0}}\right]
\end{equation}
\noindent where ${\mathcal{\ }}$ ${\mathcal{N}}_{0}$ ensures normalization to
unity in the reduced phase space $\hat{\Omega}$. Parameters $\gamma$, $\Delta P_{\phi }$, $\Delta \lambda _0$, $\Delta \lambda _1$, $P_{\phi 0}$ and $\lambda_{0}$ are, at this stage, six free parameters. 
However, it is correct to mention that Eq.~(\ref{ddf7}) is able to describe only a quite limited number of physical scenarios. Indeed, as it is shown in Refs~\citep{sonninoPRE1}-\citep{sonninoEPJD2}, Eq.~(\ref{ddf7}) correspond to stationary DDFs, written at the lowest order, obtained by adopting a very special choice of local equilibrium. In particular, this expression is inadequate for treating dynamical systems in highly anisotropic phase space. In Ref.~\citep{sonninoPRE2} we show that a large class of anisotropic dynamical systems subject to random perturbations, including particle transport in random media, can adequately be treated by assuming the validity of the MaxEnt principle of the generalized R{\'e}nyi entropies. In these cases, these entropies play the role of Liapunov functionals \citep{sonninoPRE2}.

\noindent Let us proof now that the DDF in Eq~(\ref{ddf7}) can easily be derived by applying the MaxEnt principle, subject to some restrictions, to the Shannon entropy.

\noindent The reduced phase space is defined by: $w\geq0$, $\lambda\geq0$ and $P_{\phi}\in\mathbb{R}$ . It is
invariant \ under the scaling of the variables
\begin{equation}
w\rightarrow aw\ \ ,\ \ \lambda\rightarrow b\lambda\ \ , \ \ P_{\phi}\rightarrow sP_{\phi}
\label{s0}
\end{equation}
\noindent where $a,b,s>0$ are arbitrary parameters of the dilatations group acting in $\hat{\Omega}$. We use the normalization
\begin{equation}
\int_{\hat{\Omega}}{\mathcal{F}}^{R}d{\Gamma} =1\quad {\rm where}\quad d{\Gamma} {=\left\vert {\mathcal{J}}\right\vert dwd}P_{\phi}d\lambda
\end{equation}
\noindent with ${\mathcal{J}}$ denoting the Jacobian. Since the DDF is a pseudo scalar, we have the following invariance property under the change of variables Eq.~(\ref{s0})
\begin{equation}
{\mathcal{F}}^{R}d{\Gamma}\rightarrow{\mathcal{F}}^{R}d{\Gamma}
\label{sdgamma1}
\end{equation}
\noindent The Boltzmann-Shannon entropy \citep{shannon} $\mathcal{S}(f)$ corresponding to
an arbitrary probability distribution function (PDF) $f\left(  {w,~}P_{\phi},~\lambda\right)  $ in this phase space $\hat{\Omega}$ is
\begin{equation}\label{s1}
\mathcal{S}(f)=-\int_{\hat{\Omega}}f\log f~d{\Gamma} 
\end{equation}
\noindent We can recover the form Eq.~(\ref{ddf7}) of the DDF by using the Gibbs theorem:

\noindent {\it The DDF corresponding to maximum of the entropy} (\ref{s1}) {\it with linear
restrictions generated by a set of phase space functions} $g_{0}\left(
{w,~}P_{\phi},~\lambda\right) ,\cdots,g_{N}\left(  {w,~}P_{\phi}
,~\lambda\right)$ of the form
\begin{equation}
\int_{\hat{\Omega}}fg_{k}d{\Gamma=c}_{k};~0\leq k\leq N \label{s2}
\end{equation}
\noindent {\it is given by}
\begin{equation}\label{s3}
f\left(  {w,~}P_{\phi},~\lambda\right)  =\exp\left[  -\sum\limits_{k=0}^{N}
\mu_{k}g_{k}\left(  {w,~}P_{\phi},~\lambda\right)  \right]  
\end{equation}
\noindent {\it where} $\mu_{k}$ {\it are the Lagrange multipliers that are fixed from the
constraints} Eq.~(\ref{s2}). 

\noindent Remark that the normalization Eq.~(\ref{s3}) is
assured by setting
\begin{equation}\label{s4}
g_{0}\left({w,~}P_{\phi},~\lambda\right)\equiv1 
\end{equation}
\noindent In analogy to the approach used in the previous work \citep{sonninoPRE1},
\citep{sonninoEPJD2}, where the dependence of the DDF Eq.~(\ref{ddf7}) of the
single variable $w$ was explained by using the maximal entropy principle with
scale invariant restrictions, we consider, in addition to the restriction
related to the normalization condition Eq.~(\ref{s4}) the following scale
invariant restrictions, previously used in \citep{sonninoPRE1},
\citep{sonninoEPJD2}
\begin{eqnarray}
g_{1}\left(  {w,~}P_{\phi},~\lambda\right)   &  \equiv\log(w)\label{s5}\\
g_{2}\left(  {w,~}P_{\phi},~\lambda\right)   &  \equiv w \label{s6}%
\end{eqnarray}
\noindent In order to derive the Gaussian dependence of the variable $P_{\phi}$, we use the well known homogenous functions
\begin{eqnarray}
g_{3}\left(  {w,~}P_{\phi},~\lambda\right)   &  \equiv P_{\phi}\label{s7}\\
g_{4}\left(  {w,~}P_{\phi},~\lambda\right)   &  \equiv P_{\phi}^{2} \label{s8}%
\end{eqnarray}
\noindent The $\lambda$ dependence of the density distribution function is assured by the following monomials
\begin{eqnarray}
g_{5}\left(  {w,~}P_{\phi},~\lambda\right)   &  \equiv\lambda\label{s9}\\
g_{6}\left(  {w,~}P_{\phi},~\lambda\right)   &  \equiv\lambda^{2}\label{s10}\\
g_{7}\left(  {w,~}P_{\phi},~\lambda\right)   &  \equiv w\lambda\label{s11}\\
g_{8}\left(  {w,~}P_{\phi},~\lambda\right)   &  \equiv w\lambda^{2}
\label{s12}%
\end{eqnarray}
\noindent  As shown in Ref.~\citep{sonninoEPJD2}, from the physical point of view, the {\it ad hoc} restrictions Eqs~(\ref{s7}-\ref{s12}), may be justified {\it a posteriori} by evoking the minimum entropy production theorem. According to Eq.~(\ref{sdgamma1}) and the previous form of functions
$g_{k}\left(  {w,~}P_{\phi},~\lambda\right)  $ from Eqs~(\ref{s4}-\ref{s12}),
the functional form of the linear restrictions Eq.~(\ref{s2}) remains
unchanged, only the coefficients $c_{1},...,c_{8}$ are changed
\begin{eqnarray}
c_{1}&\rightarrow& c_{1}+\log a\\
c_{2}&\rightarrow& ac_{2}\\
c_{3}&\rightarrow&sc_{3}\\
c_{4}&\rightarrow& s^{2}c_{4}\\
c_{5}&\rightarrow& bc_{5}\\
c_{6}&\rightarrow& b^{2}c_{6}\\
c_{7}&\rightarrow& abc_{7}\\
c_{8}&\rightarrow& ab^{2}c_{8}
\end{eqnarray}
\noindent From the previous formulae it follows that the linear affine sub-manifold
$\mathcal{L}$ in the functional space of probability density functions,
defined by the linear restrictions Eqs~(\ref{s2}, \ref{s4}-\ref{s12}), after
the rescaling of the variables, is translated by a constant vector. We can recover the DDF from Eq.~(\ref{ddf7}) by using Eq.~(\ref{s3}), with the functions $g_{0,...,9}\left(  {w,~}P_{\phi},~\lambda\right)  $ given by
Eqs~(\ref{s4}-\ref{s12}) and the following choice of the Lagrange multipliers
\begin{eqnarray}
\mu_{1} &=&\gamma-1\label{s13}\\
\mu_{2} &=&\frac{1}{\Theta}\left[  1+\left(  \frac{\lambda_{0}}{\Delta
\lambda_{0}}\right)  ^{2}\right]  \label{s14}\\
\mu_{3} &=&-\frac{2P_{\phi0}}{\Delta P_{\phi}^{2}}\quad;\qquad\mu_{4}=\frac{1}{\Delta
P_{\phi}^{2}}\label{s15}\\
\mu_{5} &=&\frac{-2\lambda_{0}}{\Delta\lambda_{1}\Delta\lambda_{0}}\quad ;\quad\ \!\mu
_{6}=\frac{1}{\Delta\lambda_{1}\Delta\lambda_{0}}\label{s16}\\
\mu_{7} &=&\frac{-2\lambda_{0}}{\Theta\Delta\lambda_{0}^{2}}\quad ;\qquad\ \!\mu_{8}
=\frac{1}{\Theta\Delta\lambda_{0}^{2}}\label{s17}
\end{eqnarray}
\noindent Note that the presence of the free parameter $\Delta \lambda _{1}$ in Eq.~(\ref{ddf7}) is crucial. Indeed, as we shall show in the next Section, the absence of $\Delta \lambda _{1}$ precludes the possibility of identifying the DDF, given by Eq.~(\ref{ddf7}), with the one estimated by the neoclassical theory for collisional tokamak-plasmas (see, for example, Ref.~\citep{balescu2}). In addition, it allows describing more complex physical scenarios such as, for example, the \textit{modified bi-Maxwelian} distribution function. Last and not least, in some physical circumstances, the presence of $\Delta \lambda _{1}$ is essential to ensure the normalization of the DDF.

\section{Estimation of the Free Parameters in the DDF for, Ohmic, Fully Ionized, Tokamak-Plasmas, in the weak-collisional Transport Regime}\label{estimation}
In this section, we show that it is possible to set up the free parameters appearing in the DDF in Eq.~(\ref{ddf7}) in such a way that the logarithm of the DDF coincides exactly with entropy production of tokamak-plasmas in the weak-collisional transport regime. This means that, ultimately, the DDF may be re-interpreted in terms of the Prigogine theory on the distribution functions for fluctuation of a thermodynamic variable. This leads to the attractive idea that the stationary distribution function, derived by the neoclassical theory, may also be obtained by the MaxEnt principle subject to suitable restrictions.
 
\noindent The density probability distribution of finding a state in which the values of the fluctuation of a thermodynamic variable, ${\tilde\alpha}_i$, lies between ${\tilde\alpha}_i$ and  ${\tilde\alpha}_i+d{\tilde\alpha}_i$ is (Boltzmann's constant is set to $1$) \citep{prigogine}, \citep{prigogine1}
\begin{equation}\label{i1}
{\mathcal F}={\mathcal N}_0\exp[-\Delta_I S^\alpha]\qquad (\alpha=e,i)
\end{equation}
\noindent where ${\mathcal N}_0$ ensures normalization to unity. The entropy production $\Delta_IS$ ($\geq0$) is linked to the entropy production strength $\sigma$, the thermodynamic forces $X^\mu$, and the thermodynamic flow $J_\mu$ by \citep{degroot}
\begin{equation}\label{i2}
\Delta_I S^\alpha=\int_\Omega\sigma^\alpha\ d{\bf x}\ dt=\int_\Omega\frac{n_\alpha}{\tau_\alpha}\Delta_I {\tilde S}^\alpha\ d{\bf x}\ dt=\int_\Omega X^\mu J_\mu\ d{\bf x}\ dt\geq 0
\end{equation}
\noindent Here, $n_\alpha$, $\tau_\alpha$ and $\Delta_I{\tilde S}^\alpha$ denote the number density, the relaxation time and the dimensionless entropy production (per unit particle) of species $\alpha$, respectively. $d{\bf x}$ is the spatial volume element and, in the case of Tokamak-plasmas, the integration is made over an annular shell, $\Omega$, contained between two adjacent magnetic surfaces $\psi$, $\psi+d\psi$. Integrating over time of the order of the collision time, we get $\Delta_I\hat{S}\simeq\Delta_I {\tilde S}$, with $\Delta_I\hat{S}$ denoting the entropy production per unit particle. Note that the probability density function (\ref{i2}) remains unaltered for
flux-force transformations, $X_{\kappa }\rightarrow X_{\kappa }^{\prime }$
and $J_{\kappa }\rightarrow J_{\kappa }^{\prime }$, leaving invariant the
entropy production. This property is referred to as the {\it Thermodynamic Covariance Principle} (TCP) \citep{sonninoPRE}, \citep{sonninoT}. Hence, according to Prigogine's formalism, from Eqs~(\ref{i1}) and (\ref{i2}), we see that two density distribution functions coincide if, and only if, the two expressions of $\Delta_I{\tilde S}^\alpha$ are identical {\it for all values taken by the variables}. 
 
 \noindent For easy reference, we report the main balance equations linking the DDF with entropy (i.e., the entropy production strength and the flux entropy).

\noindent $\bullet$ {\bf The equation for the entropy production strength} 
\begin{equation}\label{t7}
\sigma^\alpha=\frac{n_\alpha}{\tau_\alpha}\Delta_I{\tilde S}^\alpha=-\sum_{\beta =e,i}\int_{\mathcal V}  d{\bf v}\ [\ln {\mathcal F}^{\alpha R}({\bf v},{\bf x})]{\mathcal K}^{\alpha\beta}
\end{equation}
\noindent with ${\mathcal K}^{\alpha\beta}$ denoting the collisional operator of species $\alpha$ due to $\beta$ and ${\mathcal{V}}$ the velocity-volume in the phase-space, respectively.

\noindent $\bullet$ {\bf The equation for the flux entropy}
\begin{equation}\label{t8}
\mathbf{J}_{S}^{\alpha }(\mathbf{x})=-\int_{\mathcal{V}}d\mathbf{v}\ [\mathbf{v}-\mathbf{u}
_\alpha(\mathbf{x},{\tilde\Theta}_\alpha)]
{\mathcal{F}}^{\alpha R}(\mathbf{v},\mathbf{x})\ln {\mathcal{F}}^{\alpha R}
(\mathbf{v},\mathbf{x},{\tilde\Theta}_\alpha)=\ \frac{\mathbf{J}_{\mathcal{E}}}{T_{\alpha }}
\end{equation}
\noindent with $\mathbf{J}_{\mathcal E}$ denoting the total energy flux, and ${\bf u}^\alpha$ and $T_\alpha$ the mean velocity and temperature, respectively. The number density, the mean velocity and temperature, are provided by 
\begin{equation}\label{t9}
n_\alpha({\bf x})=\int_{\mathcal V} d{\bf v}{\mathcal F}^{\alpha R}({\bf v},{\bf x})\quad ;\quad 
n_\alpha({\bf x}) {\bf u}^\alpha({\bf x})=\int_{\mathcal V}  d{\bf v}\ {\bf v}{\mathcal F}^{\alpha R}({\bf v},{\bf x})
\end{equation}
\noindent and
\begin{equation}\label{t10}
n_{\alpha }(\mathbf{x})T_{\alpha }(\mathbf{x})=\frac{1}{2}m_{\alpha }\int_{
\mathcal{V}}d\mathbf{v}\mid \mathbf{v}-\mathbf{u}_{\alpha }\mid ^{2}{
\mathcal{F}}^{\alpha R}(\mathbf{v},\mathbf{x}) 
\end{equation}
\noindent $m_{\alpha }$ is the mass particle of specie $\alpha $. 

\noindent In this work we consider ohmic, fully ionized, tokamak-plasmas, defined as a collection of magnetically confined electrons and positively charged ions. According to the neoclassical estimation, the dimensionless entropy production of species $\alpha $, $\Delta
_{I}{\tilde S}^{\alpha }$, is derived under the sole assumption that the state of the
quiescent plasma is not too far from the reference local Maxwellian. In the \textit{local dynamical triad}, and up to the second order of the drift parameter, the dimensionless entropy production can be brought into the form (see Refs~\citep{balescu2} and \citep{sonnino})
\begin{eqnarray}\label{p12}
\!\!\!\!\!\!\!\!\Delta _{I}{\tilde S}_{neocl.}^{e}&=& \ {\tilde{\sigma}}_{\parallel }(g_{\parallel }^{(1)}-{
\bar{g}}_{\parallel }^{e(1)})^{2}+{\tilde{\kappa}}_{\parallel
}^{e}(g_{\parallel }^{e(3)}+{\bar{g}}_{\parallel }^{e(3)})^{2}+{\tilde{
\epsilon}}_{\parallel }^{e}({\bar{g}}_{\parallel }^{e(5)})^{2}\nonumber\\
\!\!\!\!\!\!\!\!&+&2{\tilde{\alpha}}_{\parallel }(g_{\parallel }^{(1)}-{\bar{g}}_{\parallel
}^{e(1)})(g_{\parallel }^{e(3)}+{\bar{g}}_{\parallel }^{e(3)})
+2{\tilde{\gamma}}_{\parallel }(g_{\parallel }^{(1)}-{\bar{g}}_{\parallel
}^{e(1)}){\bar{g}}_{\parallel }^{e(5)}+2{\tilde{\delta}}_{\parallel
}^{e}(g_{\parallel }^{e(3)}\nonumber\\
\!\!\!\!\!\!\!\!&+&{\bar{g}}_{\parallel }^{e(3)}){\bar{g}}
_{\parallel }^{e(5)} 
+{\tilde{\sigma}}_{\perp }(g_{\rho }^{(1)P})^{2}+{\tilde{\kappa}}_{\perp
}^{e}(g_{\rho }^{e(3)})^{2}-2{\tilde{\alpha}}_{\perp }g_{\rho
}^{(1)P}g_{\rho }^{e(3)}\nonumber\\
\!\!\!\!\!\!\!\!\Delta _{I}{\tilde S}_{neocl.}^{i}&=& \ {\tilde{\kappa}}_{\parallel }^{i}(g_{\parallel }^{i(3)}+
{\bar{g}}_{\parallel }^{i(3)})^{2}+{\tilde{\epsilon}}_{\parallel }^{i}({\bar{
g}}_{\parallel }^{i(5)})^{2}+2{\tilde{\delta}}_{\parallel }^{i}(g_{\parallel
}^{i(3)}+{\bar{g}}_{\parallel }^{i(3)}){\bar{g}}_{\parallel }^{i(5)}+{\tilde{
\kappa}}_{\perp }^{i}(g_{\rho }^{i(3)})^{2} 
\end{eqnarray}
\noindent Here $q_{r}^{\alpha (n)}$ [with $r=$($\rho ,\parallel ,\wedge $)]
denote the \textit{Hermitian moments of the distribution functions} and $g_{r}^{\alpha (n)},\ {\bar{g}}_{r}^{\alpha (n)}$ are the \textit{dimensionless source terms}. In particular, $g_{\parallel}^{(1)}$ and $g_{\parallel}^{\alpha(3)}$ are dimensionless parallel component of the modified electric field and of the temperature gradient of species $\alpha$, respectively. $\bar{g}_{\parallel}^{\alpha (n)}$ denote the parallel components of the additional sources, in the long mean-free-path regime (their expressions can be found in Ref.~\citep{balescu2}). Coefficients ${\tilde{\sigma}}_{r}$, ${\tilde{\alpha}}_{r}$, 
${\tilde{\kappa}}_{r}^{\alpha }$ are the dimensionless component of the 
\textit{electronic conductivity}, the \textit{thermoelectric coefficient}
and the \textit{electric} ($\alpha =e$) or \textit{ion} ($\alpha =i$) 
\textit{thermal conductivity}, respectively. Moreover, ${\tilde{\gamma}}
_{\parallel }$, ${\tilde{\delta}}_{\parallel }^{\alpha }$ and ${\tilde{
\epsilon}}_{\parallel }^{\alpha }$ are the \textit{parallel transport
coefficients} in 21M approximation. 
 
\noindent In the previous section and in Refs \citep{sonninoPRE1}, \citep{sonninoEPJD2}, we have shown that, by information theory (MaxEnt principle) or by statistical thermodynamics, it is possible to determine  the expression of the DDF, without being able, however, to fix the values of the seven parameters $\Delta P_{\phi },\Delta \lambda _{0},\Delta\lambda _{1}$ and $\Theta$. We shall see that the first three coefficients are linked to the transport coefficient whereas $\Theta$ to the source. 
 
\noindent Ultimately, Eq.~(\ref{ddf7}) represents the reference distribution function for a test particle. Hence, the logarithm of this DDF may be re-interpreted as the entropy production per unit particle (i.e., as $\Delta_I{\hat S}$). In Ref.~\citep{sonninoEPJD2} it is shown that the logarithm of the DDF, Eq.~ (\ref{ddf7}), can be brought into the form
\begin{equation}  \label{ex1}
\Delta_I{\hat S}=-(\gamma-1)\ln\Bigl(\frac{w}{\Theta}\Bigr)+\frac{w}{\Theta}+\frac{1
}{2}g_{11}{\tilde\alpha}_1^2 +\frac{1}{2}g_{22}{\tilde\alpha}_2^2+g_{12}{\tilde\alpha}_1{\tilde\alpha}_2
\end{equation}
\noindent where fluctuations ${\tilde\alpha}_\kappa$ ($\kappa=1,2$) are linked to the thermodynamic forces $X^\kappa$ and flows $J_\kappa$, by the relations \citep{onsager1}, \citep{onsager2}
\begin{equation}  \label{ex1a}
X^\kappa=\frac{\partial\Delta_I{\hat S}}{\partial{\tilde\alpha}_\kappa}\qquad;\qquad J_\kappa=\frac{d{\tilde\alpha}_\kappa}{dt}
\end{equation}
\noindent By expanding the previous expression around the reference value $w=w_0$ we obtain, up to the second order 
\begin{eqnarray}  \label{ex2}
\Delta_I{\hat S}&=&\ \Bigl[-(\gamma-1)\ln\Bigl(\frac{w_0}{\Theta}\Bigr)+\frac{w_0}{
\Theta}\Bigr]+\Bigl[-(\gamma-1)\frac{1}{w_0}+\frac{1}{\Theta}\Bigr]
(w-w_0)\nonumber\\
&+&(\gamma-1)\frac{1}{2w_0^2}(w-w_0)^2 
+\frac{1}{2}g_{11}{\tilde\alpha}_1^2+\frac{1}{2}g_{22}{\tilde\alpha}_2^2+g_{12}{\tilde\alpha}_1
{\tilde\alpha}_2+h.o.t.
\end{eqnarray}
\noindent where $h.o.t.$ stands for higher order terms. Up to a normalization constant, we have that the distribution function Eq.~(\ref{ddf7}) is approximated by a Gaussian density distribution
function (in the variable $w$) by setting to zero the coefficient of the
linear term [i.e., $-(\gamma -1)/w_{0}+1/\Theta =0$]. The \textit{global
optimality conditions} is obtained by imposing that also the sum of the
constant terms vanishes [i.e., $-(\gamma -1)\ln(w_{0}/\Theta)+w_0/\Theta =0$
]. These two requirements are simultaneously satisfied only if 
\begin{equation}  \label{ex3}
\gamma=1+E \qquad ; \qquad w_0=(\gamma-1)\Theta=E\Theta
\end{equation}
\noindent Hence, solutions~(\ref{ex3}) ensure not only a local approximation, valid up to the
second order, but also a good global approximation. In this sense the values
of $w_{0}$ and $\gamma $, provided by Eq.~(\ref{ex3}), are \textit{optimal}.
From Eqs~(\ref{ex1a}) and (\ref{ex2}) we obtain the expressions of the
thermodynamic forces 
\begin{eqnarray}  \label{ex4}
&X^1&\equiv\frac{\partial \Delta_I{\hat S}}{\partial{\tilde \alpha}_1}=g_{11}{\tilde\alpha}_1+g_{12}
{\tilde\alpha}_2  \nonumber \\
&X^2&\equiv\frac{\partial \Delta_I{\hat S}}{\partial{\tilde\alpha}_2}=g_{12}{\tilde\alpha}_1+g_{22}
{\tilde\alpha}_2 \\
&X^3&\equiv\frac{\partial \Delta_I{\hat S}}{\partial w}=\frac{1}{E\Theta^2}(w-w_0) 
\nonumber
\end{eqnarray}
\noindent By solving the previous system of equations with respect to $
\alpha_1$, $\alpha_2$ and ($w-w_0$), we find 
\begin{eqnarray}  \label{ex5}
&{\tilde\alpha}_1&={\hat g}_{22}X^1-{\hat g}_{12}X^2  \nonumber \\
&{\tilde\alpha}_2&=-{\hat g}_{12}X^1+{\hat g}_{11}X^2\quad \\
&(w-w_0)&=E\Theta^2 X^3\qquad\qquad\qquad{\rm where}  \nonumber \\
&{\hat g}_{j\kappa}&\equiv\frac{g_{j\kappa}}{g}\quad [\mathrm{with}\
(j,\kappa)=(1,2)]\qquad ; \quad g\equiv g_{11}g_{22}-g_{12}^2=\frac{1}{{\hat
g}_{11}{\hat g}_{22}-{\hat g}_{12}^2}=\frac{1}{{\hat g}}  \nonumber
\end{eqnarray}
\noindent Hence, in terms of the thermodynamic forces, the electron and ion dimensionless
entropy productions read, respectively 
\begin{eqnarray}  \label{ex6}
&\Delta_I{\hat S}^e&=\frac{1}{2}E\Theta_e^2{X_e^{3}}^2+\frac{1}{2}{\hat g}_{22}^e{%
X_e^{1}}^2+\frac{1}{2}{\hat g}_{11}^e{X_e^{2}}^2-{\hat g}%
_{12}^eX_e^{1}X_e^{2}+h.o.t. \\
&\Delta_I{\hat S}^i&=\frac{1}{2}E\Theta_i^2{X_i^{2}}^2+\frac{1}{2}{\hat g}_{11}^i{%
X_i^{1}}^2+h.o.t.  \nonumber
\end{eqnarray}
\noindent After diagonalisation, the first expression of Eq.~(\ref{ex6}) can be
brought into the form
\begin{eqnarray}\label{ex6a}
&\Delta_I{\hat S}^e&=\frac{1}{2}E\Theta_e^2{X_e^{3}}^2+\nu_1{\xi_e^1}^2+\nu_2{\xi_e^2
}^2\qquad\quad \mathrm{with} \\
&\nu_{1,2}&=\frac{1}{4}\Bigl[({\hat g}_{11}+{\hat g}_{22})\pm\sqrt{({\hat g}
_{11}-{\hat g}_{22})^2+4{\hat g}_{12}^2}\ \Bigr]  \nonumber \\
&\xi_e^1&=\frac{q_1}{h}{\hat g}_{12}X_e^{1}-\frac{q_1}{2h}({\hat g}_{11}-{
\hat g}_{22}-h)X^{2}_e  \nonumber \\
&\xi_e^2&=-\frac{q_2}{h}{\hat g}_{12}X_e^{1}+\frac{q_2}{2h}({\hat g}_{11}-{
\hat g}_{22}+h)X^{2}_e  \nonumber \\
&h&=\sqrt{({\hat g}_{11}-{\hat g}_{22})^2+4{\hat g}_{12}^2}  \nonumber \\
&q_{1,2}&=\frac{\sqrt{4{\hat g}_{12}^2+({\hat g}_{11}-{\hat g}_{22})({\hat g}
_{11}-{\hat g}_{22}\pm h)}}{\mid {\hat g}_{12}\mid\sqrt{2}}
\end{eqnarray}
\noindent Let us now consider the dimensionless electron entropy production estimated by the neoclassical theory for tokamak-plasmas in the weak-collisional transport regime \footnote{Here, calculations refer to electrons. For the ion entropy production the mathematical analysis is quite similar to the electron one. For ions, it is even simpler since the dimensionless ion entropy production is already diagonalized.}. Our aim is to find the values of the parameters such that the expression (\ref{ex6}) coincides exactly with Eqs~(\ref{p12}) {\it for all values of the thermodynamic forces}. Without considering the classical contributions, Eqs~(\ref{p12}) may be rewritten as 
\begin{eqnarray}  \label{ex7}
{\Delta_I {\tilde S}_{neocl.}^{e}}&= &\ {\tilde\epsilon}_\parallel^e{Z_e^1}^2+{
\tilde\kappa}^e_\parallel{Z_e^2}^2+{\tilde\sigma}_\parallel{Z_e^3}^2+2{
\tilde\delta}_\parallel^eZ_e^1Z_e^2+2{\tilde\gamma}_\parallel Z_e^1Z_e^3+ 2{
\tilde\alpha}_\parallel Z_e^2Z_e^3  \nonumber \\
{\Delta_I {\tilde S}_{neocl.}^{i}}&= &\ {\tilde\epsilon}^i_\parallel {Z_i^1}^2+{
\tilde\kappa}_\parallel^i{Z_i^2}^2+2{\tilde\delta}^i_\parallel Z_i^1z_i^2
\end{eqnarray}
\noindent where 
\begin{equation}  \label{ex8}
\left\{ 
\begin{array}{ll}
Z_e^1={\bar g}_\parallel^{e(5)} & \quad\mbox{$ $} \\ 
Z_e^2=g_\parallel^{e(3)}+{\bar g}_\parallel^{e(3)} & \quad\mbox{ $$} \\ 
Z_e^3\equiv g_\parallel^{(1)}-{\bar g}_\parallel^{e(1)} & \quad\mbox{ $$}
\end{array}
\right. \qquad ;\qquad\ \ \left\{ 
\begin{array}{ll}
Z_i^1=g_\parallel^{i(3)}+{\bar g}_\parallel^{i(3)} & \quad\mbox{$ $} \\ 
Z_i^2={\bar g}_\parallel^{i(5)} & \quad\mbox{ $$}
\end{array}
\right.
\end{equation}
\noindent As mentioned [see Eq.~(\ref{i2})], thermodynamic systems obtained by a transformation of forces and fluxes in such a way that the entropy production remains unaltered are thermodynamically equivalent [{\it Thermodynamic Covariance Principle} (TCP)] \citep{sonninoPRE}, \citep{sonninoT}. Hence, we should check whether the thermodynamic forces, $X^\mu$ and $Z^\mu$, are linked each other by a linear transformation\begin{equation}  \label{ex9}
\left\{ 
\begin{array}{ll}
X_e^1=Z_e^1={\bar g}_\parallel^{e(5)} & \quad\mbox{$ $} \\ 
X_e^2=Z_e^2=g_\parallel^{e(3)}+{\bar g}_\parallel^{e(3)} & \quad\mbox{ $$}
\\ 
X_e^3\equiv a_e Z_e^1 +b_e Z_e^2 +c_e Z_e^3 & \quad\mbox{ $$}
\end{array}
\right. \qquad ;\qquad\ \ \left\{ 
\begin{array}{ll}
X_i^1=Z_i^1=g_\parallel^{i(3)}+{\bar g}_\parallel^{i(3)} & \quad\mbox{$ $}\\ 
X_i^2=a_iZ_i^1+b_iZ_i^2 & \quad\mbox{ $$}
\end{array}
\right.
\end{equation}
\noindent These transformations leave invariant the entropy production i.e., $\Delta_I{\hat S}^\alpha={\Delta_I{\hat S}^{'\alpha}}$ with $\alpha=(e,i)$ \footnote{Of course, the expressions of the flows, conjugate to the thermodynamic forces $Z^\mu$, are modified in such a way that the dimensionless entropy production, $\Delta_I{\hat S}^\alpha$, remains invariant under the forces-transformation $\{X^\mu\}\rightarrow\{Z^\mu\}$.} and the two representation, $\{X^\mu\}$ and $\{Z^\mu\}$, are equivalent from the thermodynamic point of view \citep{prigogine1}, \citep{sonninoPRE}, \citep{sonninoT}. Hence, by inserting Eqs~(\ref{ex9}) into Eq.~(\ref{ex6}), we get the expressions of dimensionless entropy production (per unit particle) in terms of the thermodynamic forces $Z^\mu$. It is easily checked that the identities $\Delta_I{\hat S}^{'\alpha}\equiv {\Delta_I {\tilde S}_{neocl.}^{\alpha}}$ (with $\alpha =e,i$), are satisfied if, and only if,
\begin{eqnarray}  \label{ex11}
\!\!\!\!\!\!\!\!&a_e&=a\frac{2{\tilde\gamma}_\parallel}{\sqrt{E{\tilde\sigma}_\parallel}%
\Theta_e}\quad ;\quad b_e=\frac{2{\tilde\alpha}_\parallel }{\sqrt{E{%
\tilde\sigma}_\parallel}\Theta_e}\quad ;\quad c_e=\frac{2\sqrt{{\tilde\sigma}%
_\parallel}}{\sqrt{E}\Theta_e}  \nonumber \\
\!\!\!\!\!\!\!\! &{\hat g}^e_{11}&=\frac{2}{{\tilde\sigma}_\parallel}({\tilde\kappa}%
^e_\parallel {\tilde\sigma}_\parallel-{\tilde\alpha}_\parallel ^2)\quad
;\quad {\hat g}^e_{22}=\frac{2}{{\tilde\sigma}_\parallel}({\tilde\epsilon}%
_\parallel^e {\tilde\sigma}_\parallel-{\tilde\gamma}_\parallel ^2)\quad
;\quad {\hat g}^e_{12}=\frac{2}{{\tilde\sigma}_\parallel}({\tilde\alpha}%
_\parallel{\tilde\gamma}_\parallel-{\tilde\delta}_\parallel^e {\tilde\sigma}%
_\parallel) \\
\!\!\!\!\!\!\!\!&a_i&=\frac{2{\tilde\delta}^i_\parallel}{\sqrt{E{\tilde\kappa}_\parallel^i}%
\Theta_i}\quad;\quad b_i=\frac{2\sqrt{{\tilde\kappa}_\parallel^i}}{\sqrt{E}%
\Theta_i}\quad\ ;\quad {\hat g}^i_{11}=\frac{2}{{\tilde\kappa}_\parallel^i}({%
\tilde\epsilon}_\parallel^i {\tilde\kappa}_\parallel^i-{\tilde\delta}^i_%
\parallel{}^2)  \nonumber
\end{eqnarray}
\noindent To sum up, the set of Eqs~(\ref{ex11}) ensures that for low-collisional transport regime, the entropy production estimated by the neoclassical theory identifies with the exponent of the DDF Eq.~(\ref{ddf7}). In the neoclassical theory, plasma is heated by ohmic power. Hence, the expression of $\Theta_\alpha (X)$ can be estimated by [see Eq.~(\ref{t8})]:
\begin{equation}\label{app3}
\int_{\mathcal{V}}d\mathbf{v}\ [\mathbf{v}-\mathbf{u}
_\alpha(\mathbf{x},{\tilde\Theta}_\alpha)]
{\mathcal{F}}^{\alpha R}(\mathbf{v},\mathbf{x})\ln {\mathcal{F}}^{\alpha R}
(\mathbf{v},\mathbf{x},{\tilde\Theta}_\alpha)=\frac{\mathbf{J}_{{\mathcal{E}}_{Oh.}}}{T_{\alpha }}
\end{equation}
\noindent with $\mathbf{J}_{{\mathcal{E}}_{Oh.}}$ denoting the ohmic energy flux and [see Eqs~({\ref{t9}, \ref{t10})]
\begin{equation}  \label{app1}
\!\!\!\!\!\!\!\!\mathbf{u}_\alpha(\mathbf{x},\Theta_\alpha) =\frac{\int_{
\mathcal{V}} d\mathbf{v} \mathbf{v
}{\mathcal{F}}^{\alpha R}(\Theta_\alpha, \mathbf{v},\mathbf{x})}{\int_{
\mathcal{V}} d\mathbf{v} d
\mathbf{v} {\mathcal{F}}^{\alpha R}(\Theta_\alpha, \mathbf{v},\mathbf{x})}\quad ;\quad
T_{\alpha }(\mathbf{x})=\frac{1}{2}m_{\alpha }\frac{\int_{
\mathcal{V}}d\mathbf{v}\mid \mathbf{v}-\mathbf{u}_{\alpha }\mid ^{2}{
\mathcal{F}}^{\alpha R}(\mathbf{v},\mathbf{x})}{{\int_{
\mathcal{V}} d\mathbf{v} d
\mathbf{v} {\mathcal{F}}^{\alpha R}(\Theta_\alpha, \mathbf{v},\mathbf{x})}}
\end{equation}
\noindent In conclusion, the exponent of the DDF (\ref{ddf7}) coincides exactly with the entropy production estimated by the neoclassical theory (for tokamak-plasma in the weak-collisional regime) by setting [see Eqs~(\ref{ex6a}, \ref{ex11})]
\begin{eqnarray}\label{app1}
&\gamma&=1+E\quad ;\quad \Delta P^e_\phi=1/\sqrt{\nu_1}\quad ; \quad  \Delta\lambda^e_0\rightarrow\infty\quad ;\quad \Delta\lambda^e_1=1/\nu_2
\nonumber\\
& \xi_e^1&=P_\phi-P_{\phi 0}\quad ;\quad \xi_e^2=\lambda-\lambda_0
\end{eqnarray}
\noindent Eqs~(\ref{app1}) link the free parameters $\Delta P^e_\phi$, $\Delta\lambda^e_0$ and $\Delta\lambda^e_1$ with the transport coefficients. Note that in this case, $\Delta\lambda^e_0\rightarrow\infty$ and the presence of parameter $\Delta\lambda^e_1$ is crucial.

\noindent $\bullet$ {\bf The Local Equilibrium}

\noindent At the local equilibrium, the ${\mathcal{F}}^R_\alpha$, given in Eq.~(\ref{ddf7}), reduces to a gamma distribution function. This happens when all thermodynamic forces vanish i.e., $X^1_e=X^2_e=0$ and $X_i^1=0$. As an example of calculation, we considered L-mode, JET-plasma in the weak-collisional transport regime. By using the profiles reported in Ref.~\citep{mantica}, we estimated the electron and ion thermodynamic forces through the non-linear neoclassical theory \citep{sonninoPRE}, \citep{sonnino}. Fig.~(\ref{FDDe}) illustrates the electron forces $X^{1}_e$ and $X^{2}_e$. The intersection line corresponds to the locus of the points $(\rho,\theta)$ such that $X^{1}_e=X^{2}_e\simeq0$ (with $\rho$ and $\theta$ denoting the minor radius coordinate and the poloidal angle, respectively). Fig.~(\ref{FDDi}) shows the
curve $\rho=\rho(\theta)$ where the ion thermodynamic force, $X_i^{1}$, vanishes.

\begin{figure*}
\hfill 
\begin{minipage}[t]{.50\textwidth}
    \begin{center}  
\hspace{-1.2cm}
\resizebox{1\textwidth}{!}{%
\includegraphics{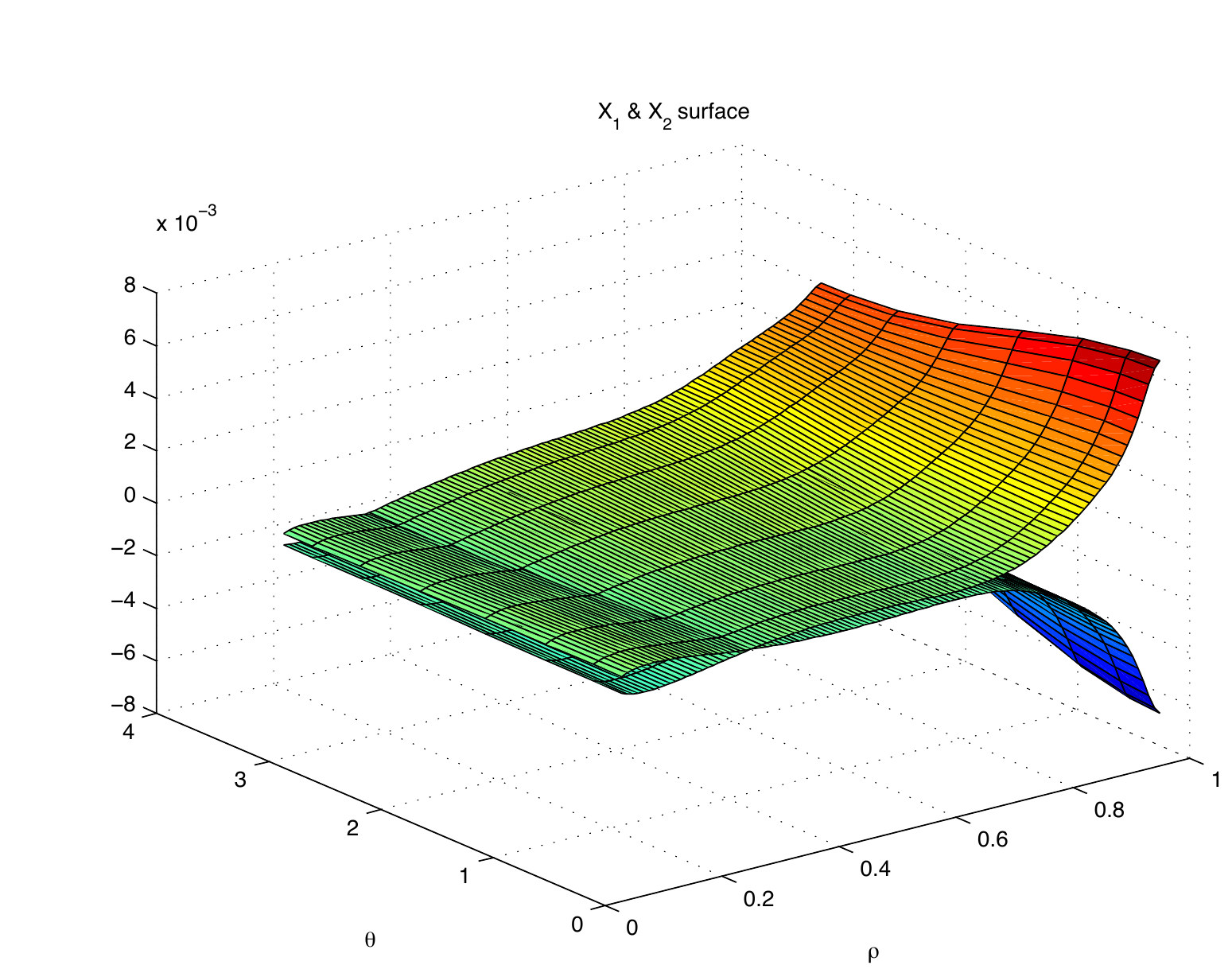}
}
\caption{ \label{FDDe} Examples of electron thermodynamic forces $X^{1}_e$ and $X^{2}_e$. These electron thermodynamic forces have been obtained by using the profiles reported in Ref.~\citep{mantica} for L-mode, JET-plasma. The intersection line corresponds to the values $(\rho,\theta)$ for which $X^{1}_e=X^{2}_e\simeq 0$ (up to the drift parameter to the square).}
\end{center}
  \end{minipage}
\hfill 
\begin{minipage}[t]{0.40\textwidth}
    \begin{center}
\hspace{-0.95cm}
\resizebox{1\textwidth}{!}{%
\includegraphics{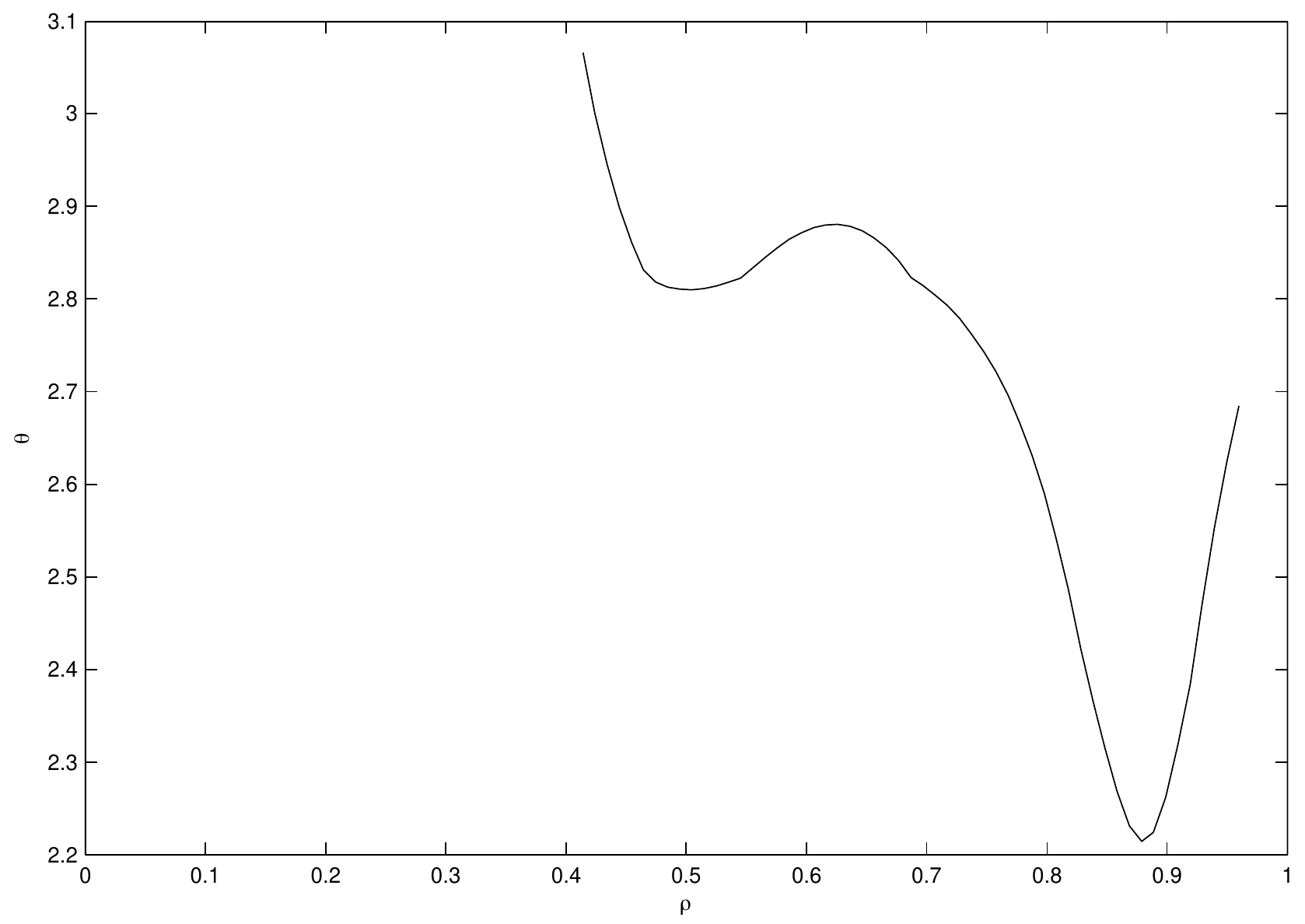}
}
\caption{Example of the locus of the points $(\rho,\theta)$ where the ion thermodynamic force vanishes ($X_i^{1}=0$). The ion thermodynamic force has been obtained by using the profiles reported in Ref.~\citep{mantica} for L-mode, JET-plasma.}
\label{FDDi}
\end{center}
  \end{minipage}
\hfill
\end{figure*}

\noindent The expressions (\ref{ddf1}) may be re-written as 
\begin{equation}\label{leq1}
P_{\phi }-\psi =\frac{B_{0}}{\Omega _{0c}}\frac{Fv_{\parallel }}{\mid B\mid}
\qquad ; \qquad \lambda-\frac{1}{2\mid B\mid}=-\frac{1}{2\mid B\mid}\frac{v^2_\parallel}{w}
\end{equation}
\noindent Note that when $P_{\phi 0}\equiv \psi$ and $\lambda_0\equiv1/(2\!\mid\!B\!\mid)$, we have that the local equilibrium is reached for $v_\parallel=0$. These conditions are useful when we analyze the weak-collisional transport regime and, in particular, the finite orbit width effects should be considered. In this case in fact, the term $(B_{0}Fv_{\parallel})/(\Omega _{0c}\!\mid\!B\!\mid)$ corresponds to the deviation of the orbit of the particle from the poloidal flux surface $P_\phi=\psi$. At $P_\phi=\psi$ we have $v_\parallel=0$  i.e., the whole parallel kinetic energy of the particle is converted into perpendicular energy, and the particle is reflected, as in a mirror.


\section{Conclusions}

We have shown that the expression of a stationary distribution function, recently proposed in literature \citep{CdT} and derived by non-equilibrium statistical mechanics \citep{sonninoPRE1}-\citep{sonninoEPJD2}, can be re-obtained solely by the information theory {\it i.e.}, by extremizing the Shannon functional, subject to {\it ad hoc} scale-invariant restrictions (MaxEnt principle). From the physical point of view, the special choice of the scale invariance restrictions can be justified by evoking the theorem of the thermodynamic of irreversible processes, such as the Minimum Entropy Production principle (MEP).

\noindent For concreteness, we analyzed the simplest case of fully ionized, L-mode, Tokamak-plasmas, in the weak collisional transport regime. We showed that it is possible to set the expression of the free parameters of the DDF obtained by MaxEnt principle, in such a way that the logarithm of the DDF identifies exactly with entropy production estimated by the neoclassical theory. In other words, the stationary DDF can be understood under the framework of Prigogine's probability theory for thermodynamic fluctuations. These results lead to the suggestive idea that the stationary DDFs used for heating Tokamak-plasmas by NBI (Neutral Beam Injection) or by ICRH (Ion Cyclotron Radio Heating) may also be re-obtained by the information theory. However, as shown in Ref.~\citep{sonninoPRE2}, Eq.~(\ref{ddf7}) is unable to treat dynamical systems in highly anisotropic phase space such as turbulent systems or anisotropic dynamical systems subject to random perturbations. Ref.~\citep{sonninoPRE2} suggests that these cases may equally be treated by the information theory. The price to pay is that the MaxEnt principle should be applied to more sophisticated expressions of entropy functionals, such as the generalized R{\'e}nyi entropies (GRE). These entropies play the role of Liapunov functionals. 

\noindent This series of works opens also another interesting perspective. Through the thermodynamical
field theory (TFT) \citep{sonninoTFT}, it is possible to estimate the DDF when
the nonlinear contributions cannot be neglected \citep{sonninopdf}. The 
task should be then to establish the link between the stationary
DDFs, derived by the MaxEnt principle applied to GRE, subject to
scale-invariant restrictions, with the ones found by the TFT. All of this will be subject of forthcoming works.

\section{Acknowledgments}
\noindent One of us (G. Sonnino) is very grateful to Dr Alessandro Cardinali of the ENEA - Fusion Department, for having actively contributed to the development of this work.

\bibliographystyle{jpp}


\bibliographystyle{jpp}


\end{document}